\documentclass[conference]{IEEEtran}
\IEEEoverridecommandlockouts
\usepackage{cite}
\usepackage{amsmath,amssymb,amsfonts}
\usepackage{algorithmic}
\usepackage{graphicx}
\usepackage{textcomp}
\usepackage{bbold}
\usepackage{xurl}
\usepackage{xcolor}
\def\BibTeX{{\rm B\kern-.05em{\sc i\kern-.025em b}\kern-.08em
    T\kern-.1667em\lower.7ex\hbox{E}\kern-.125emX}}
\begin{document}

\newcommand{\myhl}[1]{}

\title{Inferring Discussion Topics about Exploitation of Vulnerabilities from Underground Hacking Forums\\
}

\author{
\IEEEauthorblockN{Felipe Moreno-Vera\IEEEauthorrefmark{1}}
\IEEEauthorblockA{
{\IEEEauthorrefmark{1}Federal University of Rio de Janeiro (UFRJ), }
}
}

\maketitle

\begin{abstract}
The increasing sophistication of cyber threats necessitates proactive measures to identify vulnerabilities and potential exploits. Underground hacking forums serve as breeding grounds for the exchange of hacking techniques and discussions related to exploitation. In this research, we propose an innovative approach using topic modeling to analyze and uncover key themes in vulnerabilities discussed within these forums. The objective of our study is to develop a machine learning-based model that can automatically detect and classify vulnerability-related discussions in underground hacking forums. By monitoring and analyzing the content of these forums, we aim to identify emerging vulnerabilities, exploit techniques, and potential threat actors. To achieve this, we collect a large-scale dataset consisting of posts and threads from multiple underground forums. We preprocess and clean the data to ensure accuracy and reliability. Leveraging topic modeling techniques, specifically Latent Dirichlet Allocation (LDA), we uncover latent topics and their associated keywords within the dataset. This enables us to identify recurring themes and prevalent discussions related to vulnerabilities, exploits, and potential targets.


\end{abstract}


\section{Introduction}

The exploitation of vulnerabilities in the wild poses significant risks to the security and integrity of computer systems, networks, and sensitive data. With the constant evolution of cyber threats, it is crucial to identify and address vulnerabilities promptly to mitigate potential damages. Understanding how these vulnerabilities are exploited in real-world scenarios is essential for developing effective defense mechanisms and proactive security measures~\cite{basheer2021threats,campobasso2022threat}.

In recent years, there has been an increasing focus on monitoring underground hacking forums as a valuable source of intelligence regarding vulnerabilities and exploits. These forums serve as platforms for cybercriminals and hackers to exchange knowledge, discuss techniques, and share information on newly discovered vulnerabilities. By monitoring and analyzing these discussions, security researchers and practitioners can gain insights into emerging trends, exploit techniques, and potential targets in the wild~\cite{AndersonRoss,JuniperResearch}.



Understanding the exploitation of vulnerabilities in the wild through monitoring underground hacking forums can provide valuable insights for improving security practices. Early detection and proactive measures based on real-world intelligence can significantly enhance the effectiveness of vulnerability management and incident response strategies. This research aims to contribute to this growing body of knowledge by exploring the exploitation of vulnerabilities in the wild, focusing on the analysis of underground hacking forums. By leveraging machine learning and topic modeling techniques, we seek to uncover hidden patterns, emerging trends, and potential targets, providing valuable insights to strengthen cybersecurity practices and defend against evolving threats.

\textbf{Paper structure. } The remainder of this paper is structured as follows.  
In Section~\ref{sec:related}, we discuss related work, Section~\ref{sec:dataset} presents our methodology. 
Section~\ref{sec:results} reports and discusses our results, and Section~\ref{sec:conclusion} concludes.   

\section{Related Work and Background} \label{sec:related}

In what follows,  we discuss related work and background pertaining to the main themes of our work.

\subsection{Exploitation of vulnerabilities in the wild}

The exploitation of vulnerabilities in the wild is an active area of research and a growing concern in the field of cybersecurity. As attackers continuously target vulnerabilities to compromise systems and gain unauthorized access, researchers have focused on understanding these exploits and developing effective defense mechanisms. In this section, we provide a brief overview of key research contributions related to the exploitation of vulnerabilities in the wild. 

Automated vulnerability scanners, such as Nessus and OpenVAS, have been widely used to detect vulnerabilities by scanning systems and applications for known weaknesses. Additionally, manual analysis techniques, including reverse engineering and fuzzing, have been employed to uncover new and unknown vulnerabilities \cite{Liang2018FuzzingSO,Sutton2007FuzzingBF}. 

Collaboration and information sharing play a crucial role in combating vulnerabilities. Platforms and databases, such as the Common Vulnerabilities and Exposures (CVE) system and the National Vulnerability Database (NVD), provide standardized information about vulnerabilities, including their severity and available patches. Furthermore, coordinated disclosure practices, such as responsible disclosure and bug bounty programs, facilitate the reporting and fixing of vulnerabilities~\cite{Chen2016TowardsAD,Edkrantz2015PredictingVE}

\subsection{Topic modeling}

Topic modeling has been extensively researched in the field of natural language processing (NLP) and machine learning. Researchers have proposed various methods and techniques to extract latent topics from textual data, leading to advancements in understanding and organizing large document collections. In this section, we provide a brief overview of key research contributions in the field of topic modeling.

Latent Dirichlet Allocation (LDA)~\cite{Blei2001LatentDA} introduces a generative probabilistic model for topic modeling. LDA assumes that each document is a mixture of a few topics, with each topic represented as a distribution over words. Another approach is Non-negative Matrix Factorization (NMF)~\cite{Lee1999LearningTP}. This method factorizes the document-term matrix into two non-negative matrices representing the document-topic and topic-word distributions, showing good performance in topic extraction. 

Probabilistic Latent Semantic Analysis (pLSA)~\cite{Hofmann1999ProbabilisticLS} extends the latent semantic analysis by modeling documents as a mixture of latent topics using an expectation-maximization algorithm. Other approaches, such as Hierarchical Dirichlet Process (HDP)~\cite{Teh2006HierarchicalDP}, are a Bayesian nonparametric extension of LDA. Dynamic Topic Models (DTM)~\cite{Blei2006DynamicTM} capture the temporal evolution of topics in a document collection extending LDA by introducing time slices and modeling topic transitions over time.

In this work, we use topic modeling for the exploitation of vulnerabilities in underground hacking forums. We apply Latent Dirichlet Allocation (LDA) to a dataset collected from these forums to uncover hidden topics and discussions. The goal is to identify key themes related to exploit techniques, vulnerabilities, and potential targets, providing valuable insights into the landscape of vulnerability exploitation.

\begin{table}[ht!]
\centering
\setlength{\tabcolsep}{5pt}
\caption{Statistics about considered forums.  Forums are divided into boards, and boards are divided into threads. Each thread contains a list of posts. Ranked by the number of threads.}
\begin{tabular}{|l r r r r|} 
 \hline
  Forum  &  \#Users & \#Boards &
 \#Threads  &  \#Posts \\ 
 \hline
Hackforums & 630,331 & 177 & 3,966,270 & 41,571,269 \\ 
MPGH & 478,120 & 715 & 763,231 & 9,363,422 \\
Antichat & 79,769 & 60 & 242,064 & 2,449,404 \\
Offensive Community & 11,800 & 58 & 119,228 & 161,492 \\
DREADditevelidot & 44,631 & 382 & 74,098 & 294,596 \\
RaidForums & 29,038 & 73 & 33,240 & 214,856 \\
Runion & 16,719 & 19 & 16,792 & 240,632 \\
Safe Sky Hacks & 7,433 & 44 & 12,956 & 27,018 \\
The-Hub & 8,243 & 62 & 11,274 & 88,753 \\
Torum & 3,813 & 11 & 4,328 & 28,485 \\
Kernelmode Forum & 1,644 & 11 & 3,438 & 25,825 \\
Germany Ruvvy & 2,206  & 42 & 2,845 & 20,185 \\
Garage4hackers & 880 & 31 & 2,096 & 7,697 \\
Greysec & 728 & 25 & 1,630 & 9,228 \\
Stresser Forum & 777 & 16 & 702 & 7,069 \\
Envoy Forum & 362 & 76 & 454 & 2,163 \\
\hline \hline 
Total & 1,316,494 & 1,802 & 5,254,646 & 54,512,094 \\
\hline 
\end{tabular}
\label{tab:crimebb_statistics}
\end{table}

\section{Methodology} \label{sec:dataset}

In this section, we explain our methodology, present the dataset, and how to perform pre-processing.

\subsection{Datasets}

\subsubsection{CrimeBB} Cambridge Cybercrime Centre makes available sixteen underground forums through CrimeBB. 
CrimeBB comprises hierarchical data based on websites (see Figure \ref{fig:crimebb}. Table \ref{tab:crimebb_statistics} presents general statistics in CrimeBB. In CrimeBB, we have 1,316,494 users interacting on 16 websites, 1,802 boards, 5,254,646 discussion threads, and 54,512,094 posts. After analyzing our dataset, we found about 650 null values in thread titles, 7,203 posts without content, and 7,769 null usernames.

\subsubsection{NVD data} The National Vulnerability Database (NVD) is a comprehensive repository of information about software vulnerabilities and security issues. It is maintained by the National Institute of Standards and Technology (NIST) in the United States. The NVD dataset provides detailed information about known vulnerabilities in various software products, including operating systems, applications, libraries, and hardware. 

\begin{figure}[t!]
    \centering
    \includegraphics[width=\columnwidth]{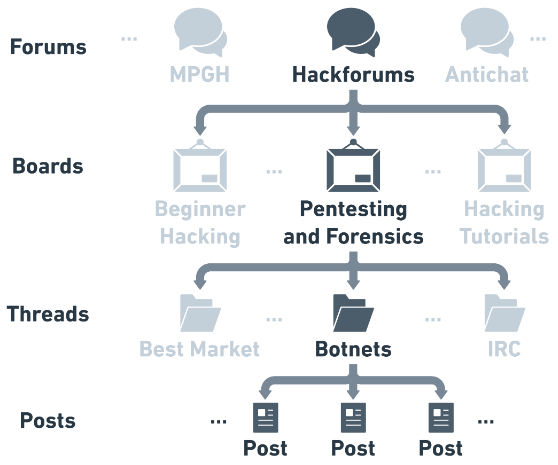}
    \caption{CrimeBB dataset, showing the hierarchical composition of websites, boards, threads, and posts.}
    \label{fig:crimebb}
\end{figure}

\subsection{Text Preprocessing}

We will divide our preprocessing into two steps: (i) nlp preprocessing, (ii)  language evaluation, and (iii) feature extraction. 

\subsubsection{NLP pre-processing}

We implement a text preprocessor that helps identify and keep relevant words to consider. We implement a library to preprocess text and evaluate language in order to facilitate our dataset preparation. We must be careful to select which word should be filtered. To do this, we filter the following characters:

\begin{itemize}
    \item \textbf{Stopwords: } These are the words in any language which does not add much meaning to a sentence. Some samples of stopwords are pronouns, adverbs, articles, etc.

    \item \textbf{Punctuations: } These are symbols that you add to a text to show the divisions between different parts of it, such as Periods, commas, semicolons, question marks, apostrophes, and parentheses.

    \item \textbf{Special Characters: } These Are the symbols used in writing, typing, etc., that represent something other than a letter (outside the 26 letters used in US English) or number, such as §, à, é, î, œ, ü, ñ, etc. 

    \item \textbf{Emojis: } These are a form of pictorial language used to express an idea, also called “digital images". These symbols denote an emotion or an action. It can be an image or textual composed symbol such as ":)", ":C", ":-)", etc.

\end{itemize}

After filtering out these characters from the raw input, we proceed to identify the language to which the input text belongs. Subsequently, we choose to convert all words using lemmatization instead of stemming. This choice is justified because, in text classification, lemmatization facilitates the creation of high-quality, contextually accurate, and semantically meaningful feature representations. This, in turn, contributes to improved classification accuracy, reduced noise, and enhanced generalization across various types of text data.\\

\begin{figure}[t!]
    \centering
    \includegraphics[width=\columnwidth]{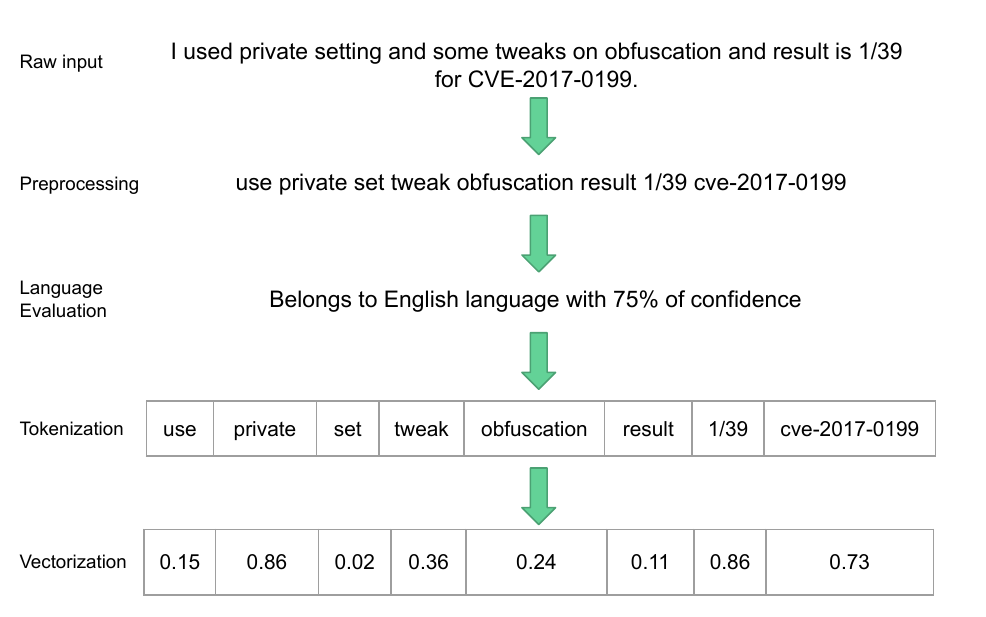}
    \caption{Text preprocessing pipeline, we show all steps from raw input until vectorization. Note that we only keep and lemmatize words to their basic root form but not all words. This post was taken from the HackForums website.}
    \label{fig:crimebb}
\end{figure}

\subsubsection{Language evaluation}

We define the \textbf{Indicator Language Function} (ILF) denoted by $\mathbb{1}_{ilf}$. In Equation \ref{eq:lang_ilf} we define our ILF, taking two parameters the word \textit{w} and the language \textit{L} to eval:

\begin{equation}
    \mathbb{1}_{ILF} (w, L) = 
    \begin{cases}
    1, & \text{if} w \in L \\
    0, & \text{Otherwise}
  \end{cases} \label{eq:lang_ilf}
\end{equation} 

Moreover, we define the \textbf{Language Ratio Function} (LRF) for a set of words and languages. This function allows us to identify and calculate what percentage of words within a phrase, paragraph, or text, in general, belongs to a specific language \textit{L}. In Equation \ref{eq:lang_lrf} we define our LRF, taking two parameters the text \textit{t} with \textit{n} words and the language \textit{L} to eval:

\begin{equation}
    \text{Ratio}_{LRF} (t, L) = \frac{1}{\text{n}} \times \sum \limits_{\substack{i=1 \\ \text{w}_i \in t}}^n \mathbb{1}_{ILF} (w_i, L) \label{eq:lang_lrf}
\end{equation} 

Finally, the \textbf{Determine Language Function} (DLF) is defined using the previously calculated Language Recognition Factor (LRF). This function determines the most probable language to which the input text belongs. In Equation \ref{eq:lang_dlf}, we define our DLF, taking a text \textit{t} as a parameter. This text is evaluated in a set of languages $\mathbb{L}$ (English and Russian language by default):

\begin{equation}
    \text{Language}_{DLF} (t) = \max\limits_{\forall L \in \mathbb{L}} \text{Ratio}_{LRF} (t, L) \label{eq:lang_dlf}
\end{equation} 

This step helps us to identify the main language within a thread (we only focus on the English language. Otherwise, we filter it). That's how We identify and filter threads from the Antichat forum. After this step, we proceed to perform text embedding, converting text into vectors. \\

\subsubsection{Tokenization \& Feature Extraction}

In this step, we perform the feature extraction from textual information. After pre-processing all textual information and performing the tokenization of words, we will use text encoding-based methods such as Bag-Of-Words (BoW) \cite{Juluru2021BagofWordsTI} and Term Frequency - Inverse Document Frequency (TF-IDF)\cite{Salton1988TermWeightingAI}. 

%


\subsection{Threads Processing}

In order to perform a Topic Modeling, we need to organize our data. In this step, we perform two main tasks: (i) Filter threads, and (ii) Label the topic of threads. 

\subsubsection{Filtering Threads}

First, we employ a filtering process to identify all posts referencing at least one Common Vulnerabilities and Exposures (CVE) code. We concatenate and merge all posts with their corresponding parent thread. Then, by using case-insensitive regular expression \texttt{cve-[0-9]\{4\}-[0-9]\{4,\}} (slightly more specific than \texttt{cve(-id)?(?i)} used  in~\cite{allodi2017economic} and~\cite{moreno2023cream}) we search for posts referring to vulnerabilities by their CVE identifiers. In Figure~\ref{fig:thread_posts_concatenation}, we present this process, showing that we ignore thread that does not cite any CVE reference. \\

\subsubsection{Labeling the Target topics}

We defined 4 main discussion topics to label the thread content. We use the following code book to label the threads manually: 

\begin{figure*}[!ht]
    \centering
   \includegraphics[width=16cm]{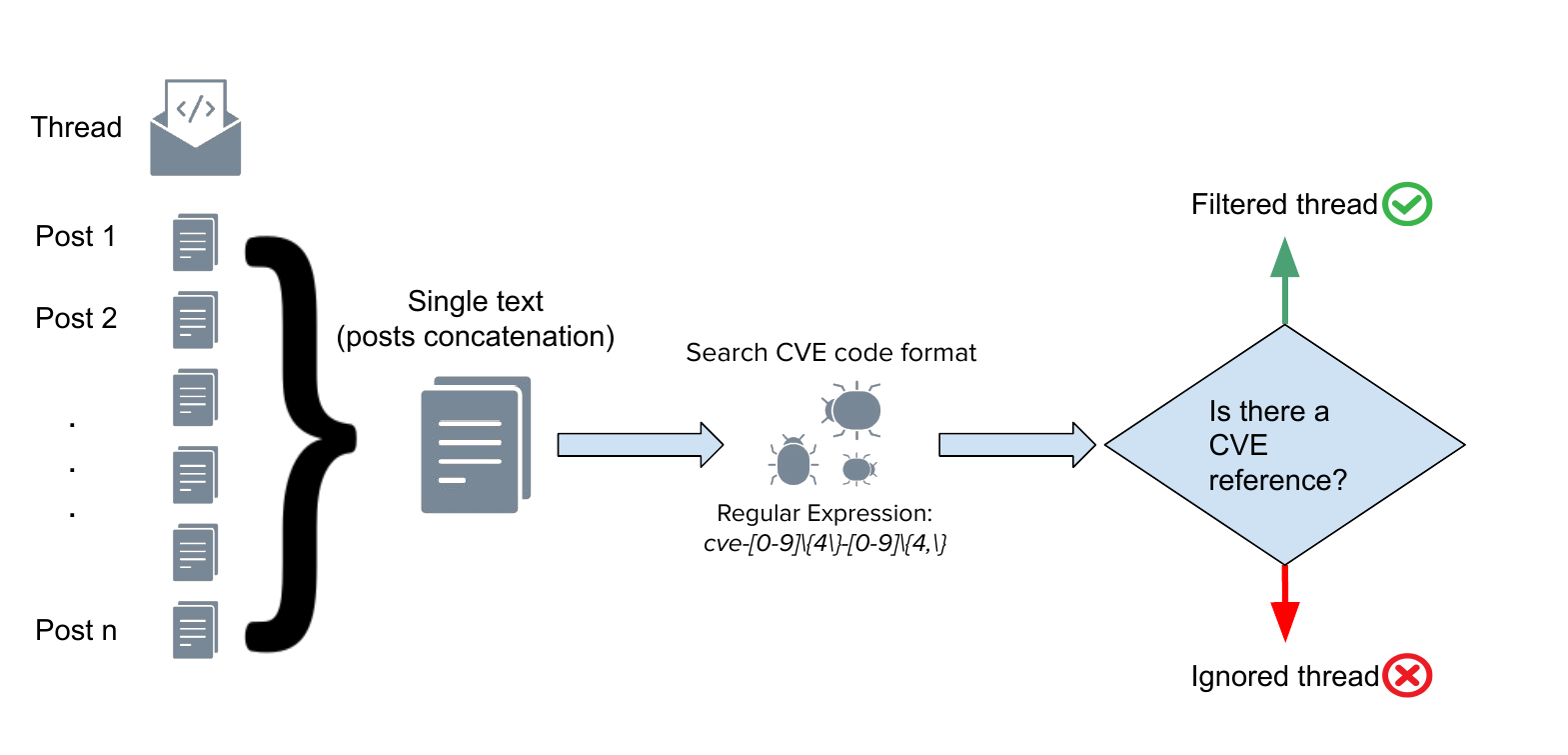}
    \caption{Post concatenation by thread: If at least one post cites a CVE code, we take all others posts from the same thread as one text sample. Otherwise, the complete thread is ignored and excluded from the dataset. This is why we don't use all labeled threads.}
    \label{fig:thread_posts_concatenation}
\end{figure*}

\begin{itemize}
    \item \textbf{PoC: } (1) contain keywords such as PoC, tutorial, guide (given the appropriate context of producing tools in a lab or controlled environment); (2) provide a tutorial description about how to build a PoC or (3) discuss vulnerabilities without signs of using exploits in the wild.   
    

    \item \textbf{Weaponization: } (1) contain keywords such as vulnerability and exploit (given the appropriate context of weaponization); (2) discuss the availability of fully functional or highly mature exploits, providing references or source code.
    

    \item \textbf{Exploitation: } (1) mention a well-known hacker group;  (2) contain references to cryptocurrencies and keywords such as bitcoin, exploitation, and attack (given the appropriate context of attacks in the wild); (3) discuss approaches to make exploits fully undetectable;  or  (4) involve markets of exploits.

    \item \textbf{Other: } Discussion about anything but the topics above.
\end{itemize}

Note: From this labeling instead of using them as labels, we decide to set them as topics.

\section{Discussions \& Results} \label{sec:results}

In this section, we discuss our analysis and the topic modeling.

\subsection{Datasets}

\subsubsection{CrimeBB} Across all CrimeBB forums, we found 4,098 posts citing 1,498 unique CVEs under 1,700 discussion threads within 149 boards. To analyze the content of threads and posts, we utilize data sourced from CrimeBB. We employ a filtering process where we specifically search for citations of CVE codes in their complete format rather than just isolated words. We found that 

\subsubsection{NVD} 
\begin{figure}[!ht]
    \centering
   \includegraphics[width=\columnwidth]{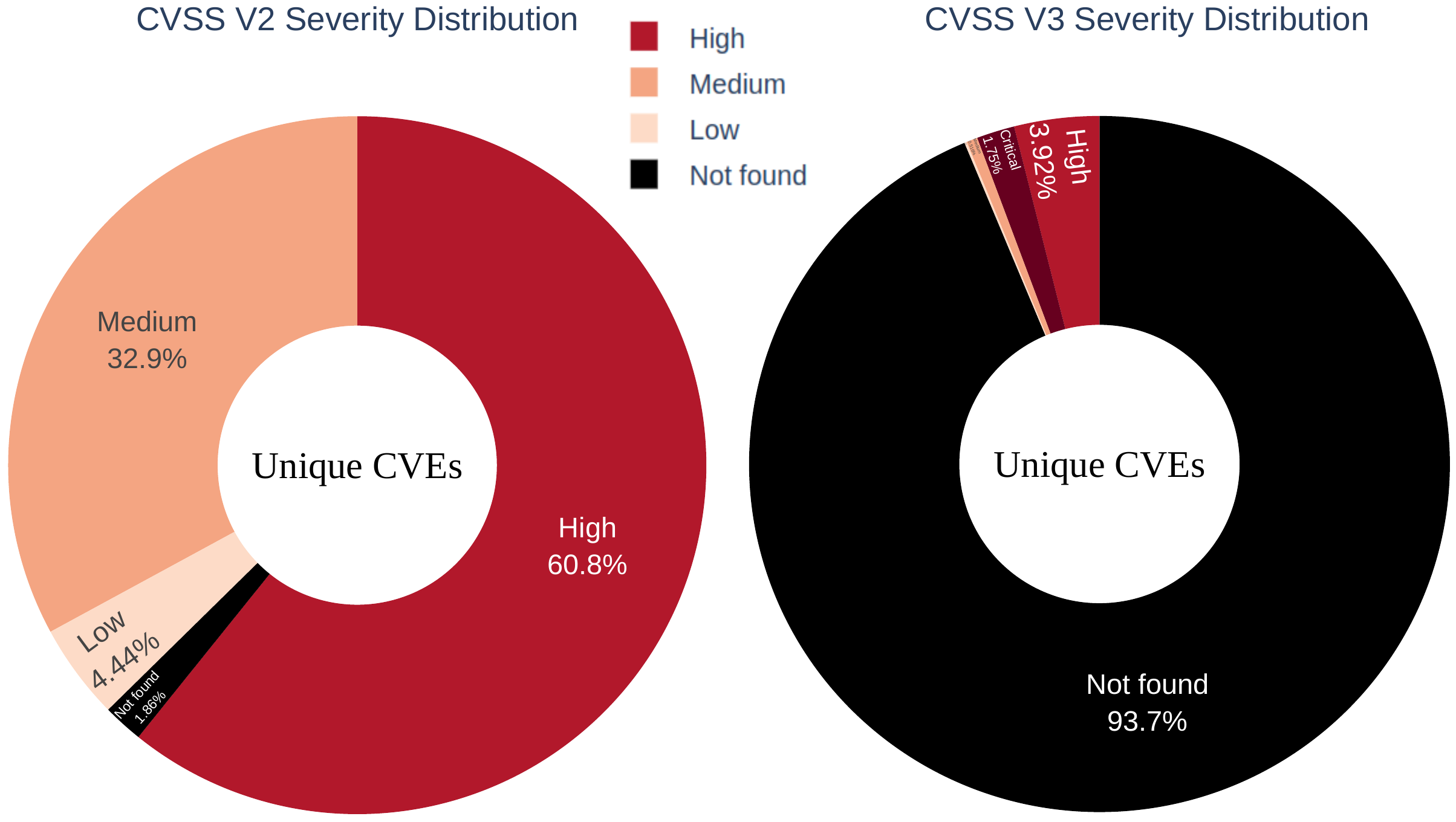}
    \caption{Common Vulnerability Scoring System (CVSS) severity level, we compare the version 2 and 3.1 of CVSS scores. We note that about 908 969 CVE codes are not found in the CVSS 3.1 version.}
    \label{fig:cvss_severity}
\end{figure}

We utilize information sourced from the National Vulnerability Database (NVD) to ascertain the characteristics of the vulnerabilities under examination. These attributes encompass factors like severity, quantified through the Common Vulnerability Scoring System (CVSS) as depicted in Figure~\ref{fig:cvss_severity}. Notably, more than 60~\% of the cited CVEs are designated with a high severity level in version 2. However, in version 3.1, we encountered difficulty in determining the present severity level of our CVE codes. 

\subsection{Text Preprocessing} 

\subsubsection{NLP pre-processing} We developed an NLP library to reproduce the text preprocessing presented in this document. The code is available at \url{https://github.com/fmorenovr/nlpToolkit/}. There, we implement an easy way to analyze languages and allow the definition of additional characters to be filtered. Besides, we found three main languages in CrimeBB: 13 websites with 4,992,945 threads discussed in English, 1 website with 242,064 threads discussed in Russian, and 2 websites with 19,637 threads discussed in Deutsch. We only focus on English websites. From this, we perform the feature extraction.

\subsubsection{Tokenization \& Feature Extraction} We perform the tokenization in a different way from previous works \cite{moreno2023cream,moreno2021understanding,moreno2019atari,leon2018car,moreno2022morarch,moreno2022comparison}; in this methodology, we will use the corpus of the language. This allows us to construct the dictionary and assign an ID to each token to identify groups of related words. We perform and join both Bag-Of-Words and TF-IDF to obtain the language corpus as features.

\subsection{Threads Processing} 

After filtering threads that cite at least one CVE code (see Figure \ref{fig:thread_posts_concatenation}), we find about 1,677 threads that explicitly cite 1,068 unique CVEs from 14 websites. Besides doing an intersection of the filtered threads and the topic-labeled threads, we got 1,067 threads with a corresponding topic. \\




\begin{figure*}[!ht]
    \centering
   \includegraphics[width=16cm]{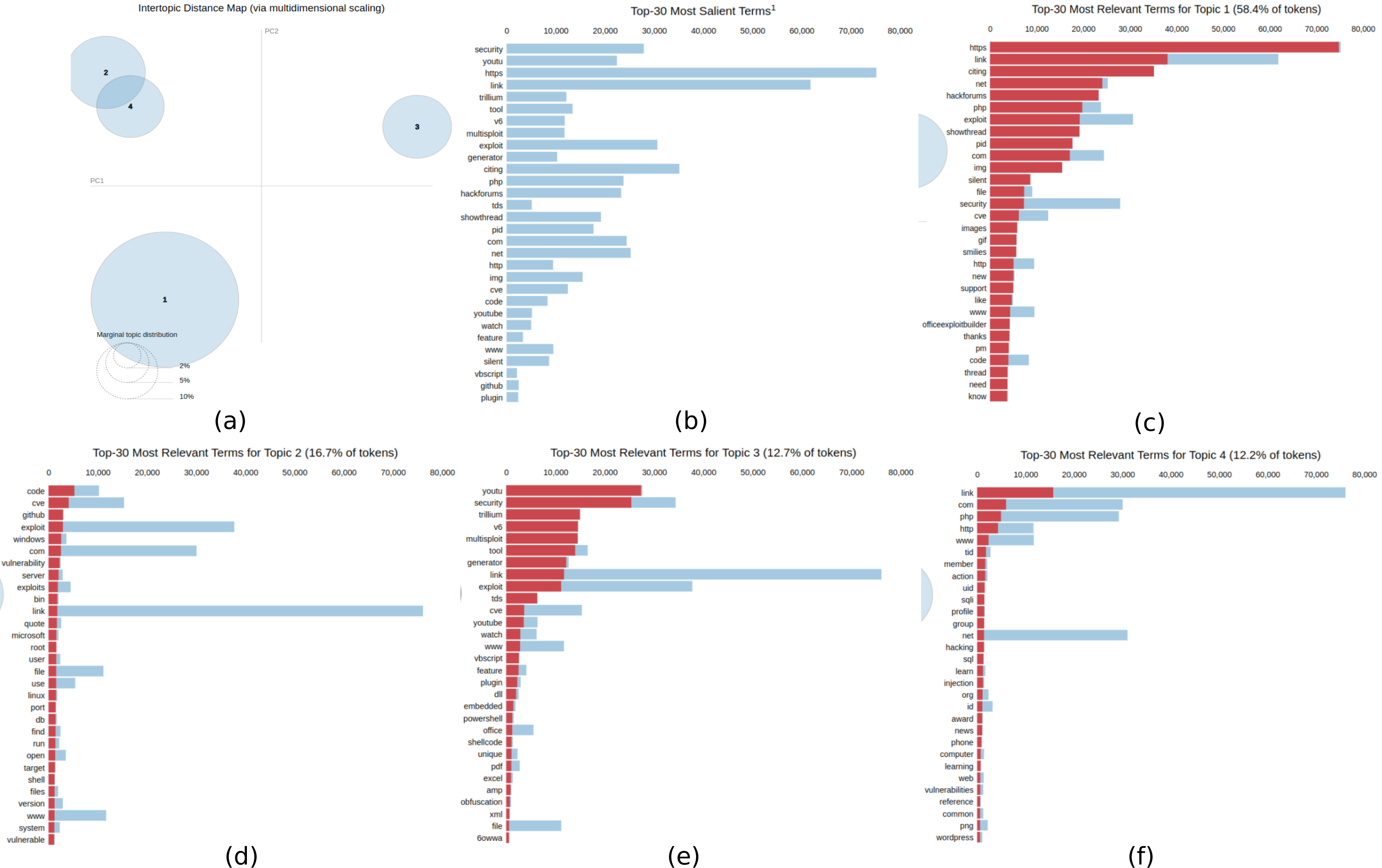}
    \caption{Topic group projection by principal components. (a) The radius of each group determines the marginal topic distribution, (b) the top 30 most salient terms, (c) the top 30 most relevant terms for topic \textit{PoC}, (d) the top 30 most relevant terms for topic \textit{Weaponization}, (e) the top 30 most relevant terms for topic \textit{Exploitation}, and (f) the top 30 most relevant terms for topic \textit{Others}.}
    \label{fig:main_terms}
\end{figure*}

\subsection{Topic modeling} 

Topic modeling is conducted on the corpus generated in the preceding stages, employing the Latent Dirichlet Allocation (LDA) algorithm. The objective is to deduce the thematic focus of the discussion by identifying pertinent terms associated with each topic. The training of the LDA model is facilitated using the Gensim library~\cite{rehurek2011gensim}. As previously indicated, our analysis encompasses four distinct topics: Proof of Concept (PoC), weaponization, exploitation, and other themes, labeled with corresponding IDs 1, 2, 3, and 4, respectively.

In Figure~\ref{fig:main_terms} (a) we show the Topic group projection by principal components, we note that topics \textit{Weaponization} and \textit{exploitation} have an intersection. This is an interesting result due to the proximity between a vulnerability from weaponization to exploitation. In Figure~\ref{fig:main_terms} (b) we show the 30 top words in all topics, we note that words such as "https", "link", "citing", and "exploit" are the most relevant.

In Figure~\ref{fig:main_terms} (c) using only the 58.4~\% of tokens the most relevant for the \textit{PoC} topic are "https", "link", "php", etc. We note that in general, those words are relevant for all topics except for "code" and "security". In Figure~\ref{fig:main_terms} (d) we show the relevant words for the topic \textit{weaponization} using only the 16.7~\% of tokens are the words "code", "cve", "github", etc. 

In Figure~\ref{fig:main_terms} (e) we show the relevant words for the topic \textit{exploitation} using only the 12.7~\% of tokens are the words "security", "trillium" (the user who cite the highest quantity of CVE codes~\cite{moreno2023cream}), "multisploit", "tool", "exploit", etc. Finally, In Figure~\ref{fig:main_terms} (f) we show the relevant words for the topic \textit{others} using only the 12.2~\% of tokens are the words "link", "com", "php", "member", "profile", "learn", etc.

We note that for each topic we have some relevant words that let us understand the main discussion. In the intersection between the topics \textit{Weaponization} and \textit{exploitation}, we have some relevant words such as "code", "cve", "exploit", "link", and "vulnerability". Besides, in the topic \textit{PoC} the most relevant word is "https", "link", "citing", "net", etc. This happens due to the nature of the discussion, sharing links, tutorials, code, etc.  From this, we observe the topic modeling could infer the discussion theme within a thread. Furthermore, we know that in each thread can be discussed several themes but only one topic.

\section{Conclusion} \label{sec:conclusion} 

In conclusion, applying topic modeling techniques to the study of exploitation in the wild offers valuable insights and benefits in the field of cybersecurity. By analyzing textual data related to vulnerabilities, exploits, and real-world attack scenarios, topic modeling contributes to a deeper understanding of the exploitation landscape. Topic modeling aids in understanding exploit trends: By extracting latent topics from data sources such as vulnerabilities, exploit forums, or security incident reports. This understanding helps security professionals, and researchers stay informed about emerging threats and prioritize their defense strategies accordingly.





\end{document}